\RequirePackage{fix-cm}
\documentclass[twocolumn]{svjour3}          
\smartqed  
\usepackage{graphicx}
\begin{document}

\title{Security checklist for IaaS cloud deployments
\thanks{We owe special thanks for J. Feket\"u for his help during the preparation of this article}
}


\author{M. H\'eder \and
        F. Bisztray \and
        Gy. Lakatos \and 
        G. Malasits \and 
        P. Ormos \and 
        E. Prunk-\'Eger \and 
        J. Rig\'o \and 
        Sz. Tenczer \and 
        E. Rig\'o
}


\institute{M. H\'eder \and
        F. Bisztray \and
        Gy. Lakatos \and 
        G. Malasits \and 
        P. Ormos \and 
        E. Prunk-\'Eger \and 
        J. Rig\'o \and 
        Sz. Tenczer \and 
        E. Rig\'o 
        \at Department of Network Security and Internet Technology \\
              MTA SZTAKI Institute for Computer Science and Control \\
              Hungarian Academy of Sciences \\
              Kende u. 13-17 \\
              Budapest \\
              Hungary, H-1111 \\
              \textit{Corresponding author:} \\ M H\'eder -- orcid.org/0000-0002-9979-9101 \\
              \email{mihaly.heder@sztaki.mta.hu}\\     
              Tel.: +35-1-279-6027\\
}

\date{Final draft date: 08 March 2016}


\maketitle

\begin{abstract}
In this article, we provide a cloud-security checklist for IaaS cloud deployments. The elements of the checklist are established by surveying the related literature on cloud-threat models and various security recommendations. We define the elements of the list on a level of abstraction that helps keep the size of the list manageable while preserving the list’s practical applicability.
\end{abstract}

\section{Introduction}
\label{intro}
Cloud computing is highly popular in all areas of business and in academia as well. Although Platform-as-a-Service and Software-as-a-Service clouds are also very popular, this paper focuses on the security issues of Infrastructure-as-a-Service (IaaS) clouds. 

IaaS cloud providers offer virtual machines (VMs) to customers with various amounts of disk storage, processing power, RAM and network access and, in many cases, backup. The users—tenants in this context—have full access to the operating system that runs in the VM. The underlying hardware is shared between tenants. 

A virtualization stack for running VMs is a mandatory part of an IaaS.\footnote{We do not count Machine-as-a-Service cloud solutions as an IaaS since they have different and very specific security characteristics.} Storage is also almost always virtualized, meaning tenants do not have direct access to the host environment’s block device or file system. Instead, most often a virtual block device is provided for them that might be backed by a file, a disk partition, a multi-node distributed-storage solution, etc. The network interface of the VMs is also virtualized. Moreover, other network components (switches, routers) might be also virtualized, implemented or supported by software-defined networking (SDN) or by other virtual network technologies. Virtualized networking nicely facilitates an IaaS service since it is more flexible in adapting to the changing demand for IaaS services than traditional networking hardware. 

Commercial IAAS providers often use the term Virtual Private Server (VPS) hosting to designate their service. This terminology refers to a difference between shared-web hosting—an older product category, also offered by many of these firms—and VM hosting. In shared-web hosting customers do not have root access to the system that produces web hosting and therefore have to accept the configuration and libraries offered; by renting a VM they can take these into their own hands.

The advantage of using an IaaS cloud for the user is that hardware and server room maintenance is not necessary. This makes the service cheaper thanks to the economies of scale, and the infrastructure can provide resources on demand on a wide-ranging scale. 

A disadvantage of IaaS from the user’s perspective is a relative loss of control over sensitive data and a general lack of transparency, as well as a decreased level of customizability of the configuration. Also from a security perspective, a big disadvantage is that users have to share the infrastructure with other parties, the number or identities of which is unknown to them. 

An attacker might want to attack the whole infrastructure or certain tenants within the infrastructure. Similarly, an attack might not just come from outside but from inside, since an attacker might buy some resources within the cloud.

IaaS cloud services might be implemented by open-source cloud software like OpenStack or OpenNebula, by various commercial products or by custom software built around a virtualization stack.

\subsection{A Note on Public vs Private Cloud}

In the cloud literature, there is a distinction made between the so-called public, private and hybrid clouds. The distinction refers to whether the cloud service is publicly offered on the market or privately used by a certain company. Hybrid clouds are on-premises private clouds that might scale out to employ publicly available resources. In this paper we do not make a distinction between these types of clouds since in the context of security there is no real difference between them. Both public and private clouds can have public IP addresses. Private clouds can have hundreds or thousands of users. Attackers might be disenfranchised employees. Even loyal employees’ personal devices might be compromised and therefore offer an attack vector to an attacker. If a cloud system is on a private network using local addresses and thus isolated from the Internet, it has an additional line of defense. However, the term “private cloud” does not imply complete privacy. In our view, limiting the cloud users to the users of a certain organization does not warrant instituting less-strict security policies. 

\subsection{The Aim and Structure of this Article}

In this article, we provide a checklist for independent or self-auditing cloud deployments with administrator access to the system. While our primary goal is not to facilitate a choice between publicly available services, some of the list items can be checked from the tenant perspective. Similarly, the main purpose of this list is not to evaluate cloud software in itself but rather actual deployments in a given hardware, network and configuration context. Still, many requirements can be derived from this list for cloud software that can be evaluated before actual deployment.

Most of the items on the checklist are cloud-security related. However, we have included some very general items that can be applied to any kind of service—for example, SELinux or AppArmor security profiles or OWASP testing—thus making the list more usable. 
The source of the items is the related literature and our experience. First, we overview the related literature; then we discuss the proposed checklist items category by category.

\section{Related Work}
\label{sec:1}
In the area of IaaS cloud security, the related work mainly discusses threat models and common security issues. Many of these works start with an examination of cloud architecture and derive cloud-specific security considerations from that. Other works provide reviews—for example \cite{Che2011} compares the NIST \cite{NIST2011}, CSA \cite{CSA2011} and Jericho forums’ cloud-threat models. 

In \cite{Chen2010}, the authors try to draw a demarcation line between general security threat types that are just merely rediscovered in the context of clouds and those that are truly cloud-specific (for any kind of cloud, including SaaS, PaaS, and IaaS). They identify the fact that users share the same resources and this increases the risk of unauthorized access by one user to another’s data, as well as the risk of an activity disruption for each other by overloading resources. Moreover, they recognize the longer trust chains that are arising from the cloud model in comparison to owned resources as a potential problem. They also address the issue of the auditability of the cloud provider. The authors also discuss nontechnical risks such as conflicts of interest between provider and consumer and the dependence of business reputation on all parties involved in the chain of trust. In our checklist, we deal with many issues arising from shared resource usage (security considerations about VMs, resource reuse, etc.), as well as with auditability (user access to logs). These considerations are built into our checklist. 

\cite{Vaquero2011} provide an IaaS cloud-threat survey. It identifies various elements of a cloud-threat model, three of which—malicious insiders, insecure interfaces and APIs, shared technology issues—are represented in our checklist. In this paper, too, threats are grouped by infrastructure elements. The paper presents an especially detailed discussion of VM hypervisors and lifecycles—some elements of our checklist in this area are based on this paper. Hypervisors in this paper are called VMMs or Virtual Machine Monitors. In the paper, Xen is used as example. 

\cite{Modi2013} present their own vulnerability model with examples. The great value in this paper for our purposes is in the examples. The paper identifies virtualization/multi-tenancy, networking, unauthorized access to management interfaces, APIs, browser problems, changes to business models, and malicious insiders as the main areas of concern.

\cite{Gonzalez2012} aim to create a cloud-security taxonomy by relying on related work as well as by examining the architecture of cloud systems. Their article is not IaaS specific. There are some general (not cloud-specific) yet still useful security considerations about the interfaces of cloud systems that appear in our checklist. The problem of shared resources appears in this paper as the isolation problem. The problem of auditing also appears; moreover, this work goes into detail about network security considerations, which are missing from many other related articles. \cite{Ali2015} present a comprehensive paper on cloud security that tries to cover every security aspect of all cloud types. This is achieved by a broad survey of related work and by deriving possible problems from different models of cloud architectures. This paper also collects the possible solutions for the problems raised. We relied on this paper for its very detailed discussion of VM isolation issues—it covers the VM rollback problem, the VM migration data leak risk, and the image reuse problem, all explained later in this paper. The article covers the general problems of virtualized networks but it does not go into similar depth with VM isolation problems. Various other works focus on the management aspects of cloud security. For instance, \cite{Rahman2015} survey incident handling practices in a cloud context, and \cite{Ramgovind2010}, besides explaining cloud types, delivery models and threat types, identify areas of security management such as regulatory compliance, data location, data segregation, recovery, investigative support, long-term viability, and data availability. 

Some works, like \cite{CSA2011} or \cite{Intel2012}, present the deployment and delivery models, as well as the main risks, but remain on a quite high abstraction level, supporting management decisions much better than cloud-security implementation.

Finally, there are security checklists that we could rely upon. The \cite{SECAAS2012} is a network-oriented, very detailed, and not just cloud-specific work that recommends IDS, VPN, logging, forensic support for hardening cloud deployments, while \cite{PCI2015} aims at defining service models and responsibilities in different cloud-delivery models in order to introduce the concept of SecaaS, that is, security as a service. We drew on both of these works for compiling our checklist; however, our approach was different. 

\section{Cloud Protection Checklist}

In this section, we provide checklist elements in different categories. The elements are either essential or preferable, depending on the seriousness of the threat posed by not meeting them. 

\subsection{Security of Web Interfaces}

Every IaaS cloud system offers a way to manage its resources through a web interface. For example, in OpenStack this interface is called Horizon, the OpenNebula counterpart of which is Sunstone. Commercial providers (e.g., Amazon) also have similar web interfaces. Since most of the resources can be managed from there, it is a high-value target. This value is even higher in the cloud systems in which the administrators of the IaaS are sharing the very same web interface with all users and thus there is no way, for example, to limit administrator access by source network (which is, by the way, an insufficient standalone security measure). This kind of risk was realized in the case of the 2013 SolusVM exploit \cite{SolusVM2013}.

Therefore, testing the security of the web interfaces of IaaS is very important. This can be done by the vendor. A checklist item for the administrators is to try to find information on whether the vendor of the cloud software has done OWASP testing \cite{OWASP2015} for the software version to be deployed. A Fuzz testing for the web API by the vendor is also recommended. 

However, even when the vendor has performed extensive testing, further local hardening is still necessary. This requirement arises from the fact that the configuration of the underlying operating system and the web server can be different from the vendor’s test configuration. This includes non-IaaS-specific web-server hardening that we do not cover in this article in detail. Some examples are hiding server and OS information from the error messages, unloading unnecessary modules, and turning off directory listing \cite{Tracy2002}. 

An essential part of web hardening is to deploy everything—both web interfaces and REST-based APIs—using SSL. Moreover, it is preferable to use multi-factor authentication for at least administrator-level access. 

The checklist items for cloud web interfaces can be found in table \ref{tab:1}.

\begin{table}
\caption{Checklist for the security of cloud web interfaces}
\label{tab:1}       
\begin{tabular}{|p{\dimexpr 0.8\linewidth-2\tabcolsep}|p{\dimexpr 0.2\linewidth-2\tabcolsep}|}
\hline\noalign{\smallskip}
Condition & Priority \\
\noalign{\smallskip}\hline\hline\noalign{\smallskip}
All webpages and REST endpoints are protected by https. & essential \\ \hline
All webpages and REST endpoints use certificates trusted by the user. & essential \\ \hline
Server-side certificates are trusted by stock browsers. & preferred \\ \hline
Client certificates are required. & preferred \\ \hline
Cloud administrators and users have separate web administration pages. & preferred \\ \hline
OWASP ASVS v3 testing is done (especially chapters V2-V8, possibly by the IaaS software vendor). & preferred \\ \hline
Fuzz testing of WEB APIs is done (possibly by the IaaS software vendor) \cite{Bounimova2013} & preferred \\ \hline
Multi-factor authentication is done, at least for users with higher privilege levels.& preferred \\
\noalign{\smallskip}\hline
\end{tabular}
\end{table}

\subsection{Security Update Management of Virtual Resources}

In IaaS systems, virtual machines are most likely to be created from templates. Virtual machine templates usually include a disk image and a specific configuration of virtual hardware and sometimes additional configuration steps executed at first boot. The virtual machine template that gets copied or ``instantiated'' when VMs are created is not a running virtual machine. While this is an advantage from the point of view of the effectiveness of the VM creation process, it is also a big security issue. That is because every modern operating system gets security (and other) updates while running. The offline template image however, will not get these updates and therefore the virtual machines created from it will contain known security vulnerabilities. 

This problem, combined with other factors such as a known vulnerability of software in certain configurations, can allow serious security incidents. The information from the template in use is available for everyone who can create a virtual machine and that might include an attacker. Moreover, in certain unfortunate network configurations (see Networking section), the attacker might monitor every VM start in the IaaS by renting just one VM in the system. This way even when the newly started VM is configured to download security patches right away after first start, an exploitable time window presents itself for the attacker. In the case of a modern OS, it can be assumed that updates happen daily or at least weekly. \textit{This basically means that even a week-old template VM can be considered as a security problem in most cases.}

A better way of handling this situation is to have ``master'' or ``etalon'' running virtual machines always with the latest security patches. This running VM can be copied when new virtual machines are started. This, however, rules out all the simple technical options for duplication. Since a running VM can write to the disk or be in the middle of an update process at any given time, making a copy of its disk can result in an inconsistent new VM. This inconsistency can be at file-system level or on the level of OS software (this can happen if the VM is copied when it is in the process of applying changes). Therefore, copy-on-write disk duplication and VM suspension for the time of duplication are not reliable solutions. Moreover, suspending the master VM each time a new VM is created does not scale --- if the VM creation happens often enough, then the update activities will not have time to finish. 
 
There are secure and technically sound alternatives. However, they require a more complex arrangement. The IaaS may monitor the processes in the master VM and make a copy --- for use as a template --- preemptively right after the automatic update activity ended within the OS. Or the IaaS may monitor a security notice feed, start an instance of the master VM when there are security updates available, wait for the update to happen in the master VM and shut down the VM. Obviously all these solutions require a way of monitoring what is happening inside the VM. This access of the IaaS to the master VM should not be replicated to the user’s VMs as this would compromise their privacy. 

Another problem presents itself if the users are allowed to take snapshots to which they can roll back later. While it is undoubtedly an additional functionality that can be very useful, it also means that users can roll back from a security-patched system to a nonpatched system. It is therefore preferred to inform the user about this risk when rolling back and also to inform the administrators about the machines rolled back. 

The checklist items related to VM security updates are summarized in table \ref{tab:2}.

\begin{table}
\caption{Checklist for security update management of virtual resources}
\label{tab:2}       
\begin{tabular}{|p{\dimexpr 0.8\linewidth-2\tabcolsep}|p{\dimexpr 0.2\linewidth-2\tabcolsep}|}
\hline\noalign{\smallskip}
Condition & Priority \\
\noalign{\smallskip}\hline\hline\noalign{\smallskip}
The template VM is configured to look for security updates as soon as possible after first launch. & essential \\ \hline
The template VM is configured to block incoming connections until security updates are done. & preferred \\ \hline
There is master VM that is updated ASAP when OS updates are available. New VMs are created replicating this master.  & preferred \\ \hline
System administrators can easily access the list of rolled back (reverted from snapshot, backup, etc.) machines. & preferred \\ \hline
The user is informed about the risks of rollback. & preferred \\
\noalign{\smallskip}\hline
\end{tabular}
\end{table}

\subsection{Security Update Management of the IaaS software}

Security update management is important also in the IaaS infrastructure itself. Updating of the infrastructure is challenging in a different way than the updating of VM images. The main challenge here is that in a typical IaaS the amount of allowed downtime is minimal, even in a scheduled fashion. The main reason for this is the large number of stakeholders concentrated in the infrastructure. The users of VMs might provide services themselves with an SLA. Therefore, very high availability is demanded from the IaaS itself. 

Security issues and updates concerning the IaaS infrastructure happen just as frequently as the software used in the VM (for instance because they use a similar Linux distribution). Yet, regular automatic updating of the IaaS is very hard to achieve because an update often requires the restart of a service. The restart of a component that provides the virtualized disk or network for VMs might disrupt the running VMs or might even corrupt their file systems. A restart of the virtualization service or the whole host means downtime for the VMs. 

Therefore, as many elements of the IaaS should be updateable separately from other elements as possible. This shows that solutions that enable live failover of components are not only essential for mitigating hardware failures but also because they enable updating. In a similar way, live migration of VMs between hosts is not only a convenience, it also allows the update of a host without the need for shutting down the VMs it runs. However, some parts of the infrastructure (e.g., the centralized disk subsystem) might only be upgraded with a complete infrastructure-wide shutdown. 

Automatic updating of the IaaS is problematic not only because of the downtime the update causes. Sometimes updates break existing functionality. While this might happen rarely, the risk of the IaaS being updated to a nonfunctional state is simply not acceptable. This means that a way is needed to test the updates, that is, a staging area is required. The IaaS software might support such staging by allowing a mixed set of versions from certain components. 

Finally, just like with operating systems, it is preferred that the IaaS updates are coming from an authentic source (e.g., signed updates) and that there are proper tools to upgrade and roll back between software versions. 

The issues around infrastructure updating most often lead to a situation in which systems containing known vulnerabilities must run for an extended period of time before they can be updated. This problem can be partially mitigated by keeping as many elements of the infrastructure as possible on an isolated network. That in turn means that every component that needs Internet access has to be able to be turned off without giving up functionality or to be configured to use a local resource (e.g., local update server, local NTP server). 

Table \ref{tab:3}. summarizes the conditions to check related to IaaS security updates.

\begin{table}
\caption{Checklist for security update management of the IaaS software}
\label{tab:3}       
\begin{tabular}{|p{\dimexpr 0.8\linewidth-2\tabcolsep}|p{\dimexpr 0.2\linewidth-2\tabcolsep}|}
\hline\noalign{\smallskip}
Condition & Priority \\
\noalign{\smallskip}\hline\hline\noalign{\smallskip}
Update-rollback tools are provided. & essential \\ \hline
Components are functional even when they have access to an isolated network only. & essential \\ \hline
Update sources are verified. & preferred \\ \hline
Failover solutions exist for every component. & preferred \\ \hline
Live migration of VMs is done. & preferred \\ \hline
Staging support is in place. & preferred \\
\noalign{\smallskip}\hline
\end{tabular}
\end{table}

\subsection{Security Considerations for Virtual Images}

The fact that many users share the same VM templates including the same image is problematic not only because of the security updates (see previous section). The images might contain information that can be exploited to attack another VM created from the same image. This kind of information can be password hashes and even salt values, software configuration details, the exact versions of every library and so on. For an attacker interested in a certain VM in the infrastructure, having access to its own instance of the same VM configuration serves a similar function to having an exact replica building for training for soldiers before they storm the actual building. 

While this problem is partially present in OS distributions in general, the setup process of an OS includes generating a unique salt value, choosing users and passwords, maybe even disk encryption, etc. This is not necessarily emulated in VM instantiation in an IaaS server.

In some IaaS systems, users can upload their VM templates into an IaaS repository or they can use a common marketplace. This introduces security problems similar to other widely used marketplaces such as Apple AppStore or Google Play. In such systems it is essential that the creator and creation time be made explicit and visible. Moreover, it is preferred that there are VM templates marked as “approved” and that the management interface can filter by creator or such tags. This is to prevent the attack in which the attacker creates a VM that looks very similar to the image the user really wants to instantiate. The user then instantiates the attacker’s image that contains a backdoor or is harmful in other ways. 

Checklist items covering virtual image security can be found in table \ref{tab:4}.

\begin{table}
\caption{Checklist for virtual image security}
\label{tab:4}       
\begin{tabular}{|p{\dimexpr 0.8\linewidth-2\tabcolsep}|p{\dimexpr 0.2\linewidth-2\tabcolsep}|}
\hline\noalign{\smallskip}
Condition & Priority \\
\noalign{\smallskip}\hline\hline\noalign{\smallskip}
VM templates do not contain any information that can be exploited by someone having access to the same VM template: 
SSH host keys should be regenerated, trusted SSH keys and custom SSL certificates should be wiped or reviewed, passwords for system and application maintenance accounts (such as debian-syst-maint MySQL admin user on dpkg based systems, or any database root password in general) should be regenerated or disabled, permissive firewall (iptables) and tcp wrappers (hosts.allow) rules should be reset. & essential \\  \hline
User-created VM templates are separate from IaaS-provided “official” templates. & essential \\  \hline
Information on creator, creation date, and verification is visible when choosing VM template. & preferred \\
\noalign{\smallskip}\hline
\end{tabular}
\end{table}

\subsection{Resource Reuse and Nullification}

After a VM is terminated, the resources belonging to it will become available for assignment to a new VM. In an IaaS, many kinds of resources store data. Different kinds of memory like RAM or possibly memory on a graphics card and different disk solutions may be available. It is essential that the new VM cannot recover any data from a previously destroyed VM. 


Some mechanisms of disk implementation are more inherently insecure in this sense than others. If the VM images are realized by LVM partitions on a block device, then the data need to be nullified or overwritten by some mechanism before the repurpose of the disk; otherwise nothing prevents the new VM from recovering data. This problem is prevented if, at the creation of the new VM, a template image overwrites the disk space. However, in an IaaS it is often possible to acquire an empty volume with a custom size. In this case, explicit nullification is essential beforehand. 

In case of copy-on-write image sharing, a single master image is used by all VMs using the image. Initially for a new VM all data is the same as in the master and, when the VM writes to the disk, the modified block is duplicated and then modified. This means that the VM cannot read from a disk area that either belongs to the master or is written over by the VM itself. 

In some cases, virtual disk blocks are not assigned continuously on the physical media. This, however, cannot be considered as a solution to the security problem because sensitive information, e.g., passwords or credit-card data, can easily be smaller than a single disk block. On the other hand, disk encryption employed by the VM does solve the problem as the content cannot be read back by another VM that does not have the encryption key. Disk encryption requires additional CPU resources though. 

Table \ref{tab:5}. contains the checklist items about resource reuse and nullification.

\begin{table}
\caption{Checklist for resource reuse and nullification}
\label{tab:5}       
\begin{tabular}{|p{\dimexpr 0.8\linewidth-2\tabcolsep}|p{\dimexpr 0.2\linewidth-2\tabcolsep}|}
\hline\noalign{\smallskip}
Condition & Priority \\
\noalign{\smallskip}\hline\hline\noalign{\smallskip}
Disk blocks are never reassigned without prior overwriting of data. & essential \\  \hline
RAM and other kinds of volatile data stores are nullified on VM start. & essential \\  \hline
Disk encryption by VM is done. & preferred \\
\noalign{\smallskip}\hline
\end{tabular}
\end{table}

\subsection{Networking}

In an IaaS, the users share the same network resources. This situation is inherently insecure as it may enable attacks that require LAN access. Active steps need to be taken even to achieve the same level of protection by isolation that is provided for those who run their own server room. 

Some IaaS networking solutions assign IP addresses from a single VLAN. In some configurations the interfaces of the virtual machines are bridged together with a physical interface of the host. In such cases, if there are no further isolation measures, this enables well-known LAN-based attacks: ARP poisoning, DHCP spoofing, CAM overflow and others. ARP traffic needs to be monitored and filtered; all DHCP server packages need to be dropped unless originated from the valid DHCP server. 

Moreover, an attacker with a VM can claim the IP address of another VM. On a simple LAN, this can cause IP conflicts leading to denial of service as well as data theft, that is, capturing IP packets not intended for the attacker. If the attacker can claim the gateway IP address, then the possibility of data theft is even bigger. Therefore, the IaaS not only needs to keep track of the assigned IP addresses, which all IaaS clouds do, but enforce that the packages intended for a given VM are not delivered to another VM. This might be done by software-defined networking (SDN). This filtering needs to work in both directions, that is, an attacker VM should not be able to send IP packages with another VM’s address as a source because this might also enable different kinds of service disruptions. 

While the drawbacks of a shared network might be mitigated, it is still preferable to run the elements of the IaaS itself in a separate network. For instance, it is not advisable to run the DHCP server, accounting, etc. components of the IaaS in a VM (just like user VMs) on the IaaS itself. 

Naturally, it is essential that the network packages that control the elements of the IaaS travel on a dedicated network, fully isolated from the network of the VMs. Otherwise it would be theoretically possible to eavesdrop or forge control messages and capture net-based virtual disk content. Also, when a virtual machine is being migrated between the hosts, at least the RAM content of the VM needs to be transmitted between hosts on network. Obviously, this needs to be an isolated network. 

Also, it is critical that every security measure should work on IPv6 with the same efficiency as it works for IPv4, if there is any IPv6 networking in the IaaS. One such measure is network address translation (DNAT/ SNAT), which is widely employed on IPv4 but has no direct counterpart on IPv6. While SNAT is helping to preserve IPv4 addresses, it also provides some security by hiding the VMs’ virtual addresses from the network. As this is not possible in IPv6, a different measure needs to be found. Also an IaaS should never automatically assign an IPv6 address to VMs alongside the IPv4 address unless that is an explicit requirement of the user. Also, it would be advisable not to mix IPv4 and IPv6 functionality within the same VM as this introduces several network access control issues.

Table \ref{tab:6}. enumerates the checklist items for cloud networking security.

\begin{table}
\caption{Checklist for cloud networking security}
\label{tab:6}       
\begin{tabular}{|p{\dimexpr 0.8\linewidth-2\tabcolsep}|p{\dimexpr 0.2\linewidth-2\tabcolsep}|}
\hline\noalign{\smallskip}
Condition & Priority \\
\noalign{\smallskip}\hline\hline\noalign{\smallskip}
Control, data networks have to be separated from VM network. & essential \\ \hline
ARP and DHCP packages must be blocked unless coming from a valid source. & essential \\ \hline
An attacker VM should not effectively claim the IP of another VM. IP packets are only allowed to be sent and received by the valid VM. & essential \\ \hline
Elements of the IaaS cannot run on the IaaS. & essential \\ \hline
VM migration happens on a separate network, isolated from the VMs. & essential \\ \hline
Security measures available for IPv4 should be supported for IPv6. & preferred \\ \hline
A VM is either providing on IPv4 or IPv6. & preferred \\ \hline
IPv6 addresses are not assigned unless explicitly requested. & preferred \\ \hline
In case of IPv6 IaaS clouds, NAT functionality is emulated by firewalling every packet except those belonging to an established flow or allowed explicitly by firewall rules. & preferred \\ \hline
Each tenant has its own VLAN. & preferred \\
\noalign{\smallskip}\hline
\end{tabular}
\end{table}

\subsection{User Access to All Security-relevant Logs and Information}

If the user runs his or her own infrastructure, he or she has access to security-relevant logs and information at every level of the architecture. In an IaaS, users cannot necessarily have access to the information on the parameters and version of the underlying virtualization container and operating system; also, they cannot determine if the software components have SELinux or AppArmor security profiles. Their access to the lower levels of the network stack for monitoring attacks might be also limited. The user, by relying on an IaaS delegate managing these layers of the system, can still access useful related information. 

Having access to this kind of information makes it possible for the users to increase their security. This includes preparedness as well as post-incident analysis and forensics abilities. While many users might opt to delegate every task related to infrastructure to the provider, this is not a reason to refrain from providing a possibility for self-monitoring or custom security enhancements. 

The corresponding checklist items can be found in table \ref{tab:7}.

\begin{table}
\caption{Checklist for log and information access}
\label{tab:7}       
\begin{tabular}{|p{\dimexpr 0.8\linewidth-2\tabcolsep}|p{\dimexpr 0.2\linewidth-2\tabcolsep}|}
\hline\noalign{\smallskip}
Condition & Priority \\
\noalign{\smallskip}\hline\hline\noalign{\smallskip}
User access is provided to view the settings of the virtualization container. & preferred \\ \hline
User access is provided for software version numbers of the underlying operating system/kernel and the virtualization architecture. & preferred \\ \hline
User access is provided to the virtualization container logs of their VMs. & preferred \\ \hline
User access is provided to network security logs concerning their VMs. & preferred \\
\noalign{\smallskip}\hline
\end{tabular}
\end{table}

\subsection{Overuse Problems}

One form of attack is causing additional costs for VM users by overusing their resources. Overuse of network quotas is quite easily caused by excessive downloading from the VM, e.g., with a botnet. High-disk IOPS, raw disk space and CPU usage might also be triggered by overusing a service that is publicly offered by the VM or even by just bloating the log files with “access denied” messages. If there are preset usage limits that stop the service when reached, then even denial-of-service attacks are possible by driving the VMs to those limits. Therefore, having a warning system in place that notifies the users about any unusual usage patterns is a security concern. For the same reason, advanced usage settings—for instance, a per-client limit for network usage—are advisable for countering attacks. Also there could be infrastructural bottlenecks (e.g., a single network node with insufficient capabilities, a common storage subsystem with low IOPS performance) that could be exploited by malicious insiders using several VMs in parallel even when VM-level resource limitations are in place.

Items to check about overuse problems can be found in table \ref{tab:8}. 

\begin{table}
\caption{Checklist for handling overuse problems}
\label{tab:8}       
\begin{tabular}{|p{\dimexpr 0.8\linewidth-2\tabcolsep}|p{\dimexpr 0.2\linewidth-2\tabcolsep}|}
\hline\noalign{\smallskip}
Condition & Priority \\
\noalign{\smallskip}\hline\hline\noalign{\smallskip}
Users get warnings on predefined usage limits. & preferred \\ \hline
Users can define usage policies such as maximum usage per time span or maximum usage per client (where applicable, e.g., based on IP addresses). & preferred \\
\noalign{\smallskip}\hline
\end{tabular}
\end{table}

\section{Conclusion}

In this article, we surveyed the literature on IaaS cloud-security checklists, guides and related articles for generally applicable IaaS hardening steps. Based on these and our own experiences with cloud hardening, we have created a checklist for IaaS cloud deployment that we believe is comprehensive yet manageable in size and realistically executable. The elements of the checklist partially reflect well-known security problems in a new shared environment and partially reflect completely new security problems introduced specifically by IaaS.


\bibliographystyle{spphys}       
\bibliography{bibfile}   

\begin{thebibliography}{10}
\providecommand{\url}[1]{{#1}}
\providecommand{\urlprefix}{URL }
\expandafter\ifx\csname urlstyle\endcsname\relax
  \providecommand{\doi}[1]{DOI \discretionary{}{}{}#1}\else
  \providecommand{\doi}{DOI \discretionary{}{}{}\begingroup
  \urlstyle{rm}\Url}\fi

\bibitem{Che2011}
J.~Che, Y.~Duan, T.~Zhang, J.~Fan, Procedia Engineering \textbf{23}, 586 (2011)

\bibitem{NIST2011}
W.~Jansen, T.~Grance, et~al., NIST special publication \textbf{800}(144), 10
  (2011)

\bibitem{CSA2011}
C.S. Alliance.
\newblock Security guidance for critical areas of focus in cloud computing
  v3.0.
\newblock \url{https://cloudsecurityalliance.org/guidance/ csaguide.v3.0.pdf}
  (2011)

\bibitem{Chen2010}
Y.~Chen, V.~Paxson, R.H. Katz, {What{\rq}s New about Cloud Computing Security?}
\newblock Tech. rep., University of California (2010)

\bibitem{Vaquero2011}
L.M. Vaquero, L.~Rodero-Merino, D.~Mor{\'a}n, Computing \textbf{91}(1), 93
  (2011)

\bibitem{Modi2013}
C.~Modi, D.~Patel, B.~Borisaniya, A.~Patel, M.~Rajarajan, The Journal of
  Supercomputing \textbf{63}(2), 561 (2013)

\bibitem{Gonzalez2012}
N.~Gonzalez, C.~Miers, F.~Redigolo, M.~Simplicio, T.~Carvalho, M.~N{\"a}slund,
  M.~Pourzandi, Journal of Cloud Computing \textbf{1}(1), 1 (2012)

\bibitem{Ali2015}
M.~Ali, S.U. Khan, A.V. Vasilakos, Information Sciences \textbf{305}, 357
  (2015)

\bibitem{Rahman2015}
R.~Ab, H.~Nurul, K.K.R. Choo, Computers \& Security \textbf{49}, 45 (2015)

\bibitem{Ramgovind2010}
S.~Ramgovind, M.M. Eloff, E.~Smith, in \emph{Information Security for South
  Africa (ISSA), 2010} (IEEE, 2010), pp. 1--7

\bibitem{Intel2012}
I.I. Center.
\newblock Planning guide: Cloud security.
\newblock \url{http://www.intel.com/content/dam/www/public
  /us/en/documents/guides/cloud-security-checklist-planning-guide.pdf} (2012)

\bibitem{SECAAS2012}
C.S. Alliance.
\newblock Cloud security alliance releases (secaas) implementation guidance
  (2012)

\bibitem{PCI2015}
P.S.S. Council.
\newblock Pci data security standard (pci dss).
\newblock \url{https://www.pcisecuritystandards.org} (2015)

\bibitem{SolusVM2013}
S.S. Team.
\newblock Important security alert – all solusvm versions.
\newblock
  \url{http://solusvm.com/blog/important-security-alert-all-solusvm-versions/}
  (2013)

\bibitem{OWASP2015}
O.W.A.S. Project.
\newblock Application security verification standard 3.0.
\newblock \url{https://www.owasp.org/images/6/67/
  OWASPApplicationSecurityVerificationStandard3.0.pdf} (2013)

\bibitem{Tracy2002}
M.~Tracy, W.~Jansen, M.~McLarnon, NIST Special Publication \textbf{800}, 44
  (2002)

\bibitem{Bounimova2013}
E.~Bounimova, P.~Godefroid, D.~Molnar, in \emph{Proceedings of the 2013
  International Conference on Software Engineering} (IEEE Press, 2013), pp.
  122--131

\end{thebibliography}

%
%

\end{document}